\documentstyle[12pt]{article}
\begin{document} 
\thispagestyle{empty} 
\begin{flushright}
UA/NPPS-15-2001\\

\end{flushright}
\begin{center}
{\large{\bf THE PATH FROM CHEMICAL TO THERMAL FREEZE-OUT \\}} 
\vspace{2cm} 
{\large A. S. Kapoyannis}\\ 
\smallskip 
{\it University of Athens, Division of Nuclear and Particle Physics,\\ 
GR-15771 Athens, Greece}\\ 
\vspace{1cm}

\end{center}
\vspace{0.5cm}
\begin{abstract}
The evolution of a hadronic system after its chemical decomposition is
described through a model that conserves the hadronic multiplicities to
their values at chemical freeze-out. The state of the system is found as
function of temperature and the corresponding baryon density is evaluated.
The baryon density at thermal decoupling is also computed.
\end{abstract}

\vspace{7cm}


\newpage
\setcounter{page}{1}

{\large \bf 1. Introduction}

Thermal approaches have extensively been used to describe the particle
multiplicities which emerge from high energy collisions [1-16]. The results of
such approaches are satisfactory since they are able to predict a large number
of different experimentally measured hadronic multiplicities as functions of
a few thermodynamic variables, such as temperature, volume and chemical
potentials.

The extracted parameters from such approaches are associated with ``chemical
freeze-out'', i.e. the point where the chemical composition of the fireball
is fixed. After this stage the particles continue to interact until their
momentum distribution is fixed, as well. This second point is called
``thermal freeze-out''.

Since a set of experimentally measured abundances {\it define} the chemical
freeze-out point these abundances have to remain fixed through the whole
cooling process until thermal freeze-out. After all, the particles are
measured once freeze-out has been completed.

Various authors have used models of thermally equilibrated relativistic
hadronic abundances to determine the chemical freeze-out parameters. In some
of these models the hadrons are non-interacting particles [2-7] and in others
a kind of interaction among them has been included [1,8-16]. In this paper
the main focus will be on a model of a non-interacting hadronic gas formulated
in the grand canonical ensemble, called ``Ideal Hadron Gas'' (IHG) [2-5]. It
will be argued that in this model, as
in the rest of thermal models, it is {\it not} possible to fulfil the
requirement that all the particle multiplicities will remain fixed for the
whole way from chemical decoupling to thermal freeze-out. Considering for
example IHG, the partition function is expressed as function of
$(V,T,\{\lambda\})$ where $\{\lambda\}$ are fugacities associated with
quantum numbers, such as Baryon Number, $B$, Charge, $Q$ and Strangeness, $S$,
as well as the departure from absolute chemical equilibrium. In this
particular case the total number of these fugacities
is limited to at most five. Not all of them are independent since
quantities such as $<B>$, $<Q>$, $<S>$ have to remain fixed, limiting the
total number of independent parameters to four.
The temperature at thermal
freeze-out point is generally different from the chemical freeze-out one.
So it will be impossible for someone who will use the same model at thermal
freeze-out point to have all the multiplicities fixed to their chemical
freeze-out values. The existing free parameters will not be enough.

In this work the necessity to have fixed particle numbers will be used to
construct a thermal model which will determine the evolution of the hadronic
system after its chemical freeze-out. With the use of this model the
construction of the ``path'' followed by the system in the diagram of
temperature as function of baryon density will be possible.

\vspace{0.3cm}
{\large \bf 2. The model and its application}

In the context of IHG the grand canonical partition function, formulated
in the Boltzmann approximation, has the form
\begin{equation}
\ln Z(V,T,\{\lambda\})_{IHG} = V \sum_{i} \lambda_{QN_i} \sum_j Z_{H_{ij}}(T)
\equiv V \sum_{i} \lambda_{QN_i} \sum_{j} \frac{T}{2\pi^2} g_{ij} m_{ij}^2
K_2(\frac{m_{ij}}{T})\;,
\end{equation}
where $i$ runs over all hadronic families such as mesons, $N$ Baryons,
$\Lambda$ Baryons, etc. and $j$ represents the specific member of the
family with degeneracy factor $g_{ij}$ and mass $m_{ij}$. $\lambda_{QN_i}$
stands for the product of all the fugacities associated with the particular
family. These fugacities can either be quantum numbers fugacities related to
Baryon number, Strangeness, etc. or to quark flavour\footnote{For example,
for $\Xi^-$ Baryons, $\lambda_{QN}$ would read
$\lambda_B\lambda_Q^{-1}\lambda_S^{-2}\gamma_s^2$ or
$\lambda_d\lambda_s^2\gamma_s^2$. One can look for example in [16], eq. 14,
to find out how the two sets of fugacities are related.}.

Using the above relation one can evaluate particle abundances if one extends
the partition function by introducing a fugacity $\lambda_{ij}$ for every
particle. After calculating the particle number one has to set
$\lambda_{ij}=1$ [17], so again the particle number is only expressed as
function of the quantum numbers fugacities.

Now, if someone wishes to keep the particle numbers fixed, e.g. at their
chemical freeze-out values,
it is only natural to use the fugacities $\lambda_{ij}$, but
with the difference that they are allowed to be $\lambda_{ij} \neq 1$.
This model is called Fixed Particle Numbers (FPN) model and
accordingly the partition function depends on $\lambda_{ij}$'s
\begin{equation}
\ln Z(V,T,\{\lambda\})_{FPN} = V \sum_{ij} \lambda_{H_{ij}} Z_{H_{ij}}(T)
\equiv V \sum_{ij} \lambda_{H_{ij}} \frac{T}{2\pi^2} g_{ij} m_{ij}^2
K_2(\frac{m_{ij}}{T})\;,
\end{equation}
where $\lambda_{H_{ij}}$ is product of quantum numbers as well as particle
number fugacities\footnote{For example,
for $\Xi(1530)^-$ Baryon, $\lambda_{H}$ would read
$\lambda_B\lambda_Q^{-1}\lambda_S^{-2}\gamma_s^2\lambda_{\Xi(1530)^-}$.}.
The mean particle number can be evaluated through the
relation
\begin{equation}
<N_{ij}>=\left.\lambda_{ij}
\frac{\partial \ln Z(V,T,\{\lambda\})_{FPN}}{\partial\lambda_{ij}}
\right|_{\{\lambda\}\neq\lambda_{ij}}\;,
\end{equation}
where $\{\lambda\}\neq\lambda_{ij}$ means that all fugacities except
$\lambda_{ij}$ are considered as constants.

A thermal model, like IHG, can then be used to extract the set of chemical
freeze-out parameters $\{V,T,\lambda_B,\lambda_S,\cdots\}$ which best fit a
set of experimentally measured multiplicities. With these given parameters the
numbers $<N_{ij}>$ of all particles that compose the hadronic gas can be
calculated. All these numbers have to stay {\it fixed}  during the evolution
of the system after chemical freeze-out, so $\lambda_{ij}$ will be allowed
to become different than one. With the use of (3) this
requirement takes the form\footnote{The primed variables in this paper will
generally be related to subsequent points of the chemical freeze-out point.}
\[
<N_{ij}>=<N_{ij}'>\Leftrightarrow
V\lambda_{QN_i}Z_{H_{ij}}(T) =
V'\lambda_{H_{ij}}Z_{H_{ij}}(T')\Leftrightarrow
\]
\begin{equation}
\lambda_{H_{ij}}=
\frac{V\lambda_{QN_i}Z_{H_{ij}}(T)}{V'Z_{H_{ij}}(T')}\;.
\end{equation}
In the last equation $\lambda_{H_{ij}}$ may contain quantum numbers fugacities
as well the particle number fugacity. As it is shown in the Appendix it is
not possible to evaluate each quantum number fugacity, but this is irrelevant
since the full product of fugacities can be calculated. Let me point out that
all quantum numbers are automatically conserved because they are linear
combination of the particle numbers.

The volume of the system at chemical freeze-out, $V$, on the other hand,
need not necessarily stay fixed. If $V'$ at temperature $T'$ is different
from $V$, then all fugacities given by (4) depend on a multiplicand factor
$\frac{V}{V'}$, which cannot be determined from the constraints imposed by
the conservation of particle numbers. An additional constraint has to be
applied. For example conservation of entropy can be assumed\footnote{A lot
of authors assume isentropic evolution of the system, e.g. see [19].}.

The entropy of the system can be calculated from\footnote{The symbol of
entropy is tilded in order not to be confused with the symbol of Strangeness.
$K$ can be set equal to one.}
\begin{equation}
\tilde{S}=-\left(\frac{\partial[-T\ln Z(V,T,\{\mu\})]}{\partial T}
\right)_{V,\{\mu\}}\;,
\end{equation}
where $\mu$ represents the chemical potential associated with fugacity
$\lambda=\exp(\mu/T)$. Applying (5) to the partition function
(2)\footnote{The IHG partition function (1), where $\lambda_{ij}=1$, can be
used for the evaluation of the entropy at the chemical freeze-out point.} the
constraint of fixed entropy will read
\[
\tilde{S}=\tilde{S}'\Leftrightarrow
\]
\[
\Leftrightarrow \ln Z(V,T,\{\mu\})+
VT \sum_{ij} \lambda_{QN_i} \frac{\partial Z_{H_{ij}}(T)}{\partial T}-
VT \sum_{ij} \lambda_{QN_i} \frac{\mu_{QN_i}}{T^2} Z_{H_{ij}}(T)=
\]
\[
=\ln Z(V',T',\{\mu'\})+
V'T' \sum_{ij} \lambda_{H_{ij}} \frac{\partial Z_{H_{ij}}(T')}{\partial T'}-
V'T' \sum_{ij} \lambda_{H_{ij}}\frac{\mu_{H_{ij}}}{T'^2} Z_{H_{ij}}(T')\;.
\]
With the use of (4) the last equation becomes
\[
VT \sum_{ij} \lambda_{QN_i} \frac{\partial Z_{H_{ij}}(T)}{\partial T}-
V \sum_{ij} \lambda_{QN_i} \ln(\lambda_{QN_i}) Z_{H_{ij}}(T)=
\hspace{7cm}
\]
\[
=V'T' \sum_{ij} \frac{V\lambda_{QN_i}Z_{H_{ij}}(T)}{V'Z_{H_{ij}}(T')}
\frac{\partial Z_{H_{ij}}(T')}{\partial T'}-
V' \sum_{ij} Z_{H_{ij}}(T')
\frac{V\lambda_{QN_i}Z_{H_{ij}}(T)}{V'Z_{H_{ij}}(T')}
\ln(\frac{V\lambda_{QN_i}Z_{H_{ij}}(T)}{V'Z_{H_{ij}}(T')})\Leftrightarrow
\]
\[
\Leftrightarrow
T \sum_{ij} \lambda_{QN_i} \frac{\partial Z_{H_{ij}}(T)}{\partial T}-
\sum_{ij} \lambda_{QN_i} \ln(\lambda_{QN_i}) Z_{H_{ij}}(T) =
\hspace{7cm}
\]
\begin{equation}
\hspace{3cm}
=T' \sum_{ij} \lambda_{QN_i} \frac{Z_{H_{ij}}(T)}{Z_{H_{ij}}(T')}
\frac{\partial Z_{H_{ij}}(T')}{\partial T'}-
\sum_{ij} \lambda_{QN_i} Z_{H_{ij}}(T) 
\ln(\frac{V\lambda_{QN_i} Z_{H_{ij}}(T)}{V'Z_{H_{ij}}(T')}) \;.
\end{equation}
Setting $x\equiv\frac{V'}{V}$, (6) can be solved for $x$ to
give\footnote{All hadrons with masses up to 2400 MeV are included in the
calculations corresponding to FPN and to IHG models.}
\begin{equation}
x=\exp\left[\frac
{\sum_{ij} \lambda_{QN_i} Z_{H_{ij}}(T) 
\ln(\frac{Z_{H_{ij}}(T)}{Z_{H_{ij}}(T')})+
T \sum_{ij} \lambda_{QN_i} \frac{\partial Z_{H_{ij}}(T)}{\partial T}-
T' \sum_{ij} \lambda_{QN_i} \frac{Z_{H_{ij}}(T)}{Z_{H_{ij}}(T')}
\frac{\partial Z_{H_{ij}}(T')}{\partial T'}}
{\sum_{ij} \lambda_{QN_i} Z_{H_{ij}}(T)}
\right]\;.
\end{equation}

Equation (7) can be used to evaluate the volume expansion ratio as the
system has cooled to a temperature $T'$ less than the chemical freeze-out
temperature $T$. With the use of the same equation, quantities like the
baryon density of the system can be calculated at $T'$. One has to remember
that baryon number is also fixed with the imposition of the constraints (4).
So
\begin{equation}
n_B=\frac{<B'>}{V'}=\frac{<B>^{ch}}{V'}=\frac{V}{V'}\cdot\frac{<B>^{ch}}{V}=
\frac{n_B^{ch}}{x}\;.
\end{equation}

Other constraints which have to be applied to the system and are connected
to quantum numbers, like
\begin{equation}
<S>=0\;,\hspace{1cm}
\frac{<B>}{<Q>}=\left(\frac{<B>}{<Q>}\right)^{ch}\;,\hspace{1cm}
<|S|>=<|S|>^{ch}\;,
\end{equation}
are also satisfied.

Thus, the contour which is followed by the system after the fixation of its
chemical composition until its thermal freeze-out can be evaluated. This
contour can be defined on a $(T,n_B)$ plane with the use of eqs. (7) and (8)
but not on a $(T,\mu_B)$ plane for the reasons explained in the Appendix.

As an example the model is used to depict the path followed by hadronic
systems which have been formed at different interactions at SPS after their
chemical freeze-out. The chemical freeze-out parameters used, along with the
corresponding references are listed in Table 1. From a variety of thermal
analyses performed by different authors the particular ones have been chosen
because they allow for partial strangeness equilibrium ($\gamma_s\neq1$) and
they use most recent available values for the experimentally measured
hadronic multiplicities. The values of Table 1 are then used, for each
interaction separately, as input to the equations $<S>=0$ and
$\frac{<B>}{2<Q>}=\beta$,\footnote{$\beta$ is fixed from the baryon number
and charge of the participant nucleons, e.g. see [16].} evaluated through IHG,
to determine the rest of the fugacities.
Thus the whole set of chemical freeze-out parameters
$(T,\mu_B,\mu_Q,\mu_S,\gamma_s)$ are calculated and also the products of
fugacities $\lambda_{QN_i}$ in (1) are also set.

Giving different values to temperature $T$, equation (8) can be used to
calculate the corresponding baryon density. The resulting paths for $S+S$,
$S+Ag$ and $Pb+Pb$ interactions are shown in Figure 1. For the $Pb+Pb$
interaction the thermal freeze-out temperature is calculated in Refs. [21] and
[22]. For these values baryon density at thermal freeze-out $n_B^{ther}$ can
be evaluated. The results are listed in Table 2. The path for $Pb+Pb$ is
followed until the lower temperature (of the two given in Refs. [21,22]) is
reached.

In order to compare FPN with an IHG model which presents the closest
characteristics with it, points that are subjected to the
constraint that the baryon number and the entropy is fixed ($<B>=<B>^{ch}$
and $\tilde{S}=\tilde{S}^{ch}$)\footnote{The rest of the constraints on quantum
numbers, like $<S>=0$, etc., are applied as well.} are also depicted on
Figure 1. Let me emphasise that at these points
the particle numbers are not conserved. But the IHG points have
no problem to be depicted on a ($T,\mu_B$) plane. This is done in Figure 2.
On this Figure there is also depicted an ``equivalent'' value of baryon
chemical potential ${\mu_B}_{eq}$ for FPN model as function of temperature.
This chemical potential is calculated through IHG and the only connection
it has with FPN is that it gives the same baryon density
\begin{equation}
n_B(T,{\mu_B}_{eq})_{IHG}=n_B(T)_{FPN}\;.
\end{equation}
For comparison with FPN, points that correspond to calculations through IHG
for the given thermal freeze-out temperatures of Table 2 have also been
depicted on Figs 1 and 2.

Finally in Figure 3 the ratio $x=V'/V$, where $V$ is the chemical freeze-out
volume, is plotted as function of temperature for FPN and for IHG (with fixed
entropy and baryon number).

\vspace{0.3cm}
{\large \bf 3. Conclusion}

After chemical freeze-out the collisions among hadrons that compose the
hadronic gas can no longer change its chemical composition. Following this
requirement an ideal hadron gas model (FPN) has been presented that keeps the
multiplicity of every particle fixed to the value dictated by the chemical
freeze-out conditions. In the context of FPN the constraints of conservation
of quantum numbers are broken up to a larger number of constraints, these of
conservation of particle numbers. The chemical potentials of quantum numbers
are no longer ``good'' variables to describe the evolution of the system. Of
course the fugacities of particle numbers used as variables in FPN are not
``free'' parameters. Their values are fixed from the given set of the quantum
numbers fugacities at chemical freeze-out. So the evolution of a hadronic
system is described as function of temperature and baryon density (after
imposing conservation of entropy). This is done for three SPS interactions.

Following this evolution and using values of thermal freeze-out temperature
extracted for the $Pb+Pb$ interaction the baryon density at freeze-out is
evaluated. As the temperature at thermal decoupling for various interactions
can be calculated using transverse mass spectra or HBT analysis [23] the same
procedure can be applied to evaluate the corresponding baryon density before
free streaming for these interactions. Finally let me point out that the
necessity to keep the particle multiplicities fixed after chemical freeze-out
can be fulfilled to any thermal model, apart from IHG, through the use of
the particle fugacities.

\vspace{0.3cm}
{\large \bf Acknowledgement}
I would like to thank Professor N. G. Antoniou and Professor C. N. Ktorides
for reviewing the manuscript and for useful remarks.

\vspace{0.3cm}
{\large \bf Appendix}

It will be argued that an ambiguity presents itself when someone tries
to evaluate the quantum numbers fugacities.
Let us suppose that the system is initially in a state which is
determined by volume $V$, temperature $T$, quantum fugacities
$\lambda_B,\lambda_S,\cdots$ and particle number fugacities $\lambda_{ij}$.
The question that arises is whether it is possible to determine the
thermodynamic variables connected to a subsequent temperature $T'$.
Normally the new fugacities could be calculated from a set of $n$ equations
of the form\footnote{The primed quantities are connected to $T'$.}
\[
<B(V,T,\{\lambda\})>=<B'(V',T',\{\lambda\}')>\;,
\]
\begin{equation}
<S(V,T,\{\lambda\})>=<S'(V',T',\{\lambda\}')>\;,\cdots\;
\end{equation}
which insure for the conservation of the $n$ quantum numbers and a set of $m$
equations of the form
\begin{equation}
<N_{ij}(V,T,\{\lambda\})>=<N_{ij}'(V',T',\{\lambda\}')>\;,
\cdots\;
\end{equation}
which insure for the conservation of the number of the $m$ particle species
that are available in the hadronic gas. If one tries to solve the above
set of the $n+m$ equations one will find out that it is impossible to
determine all the fugacities. The reason for this is that the $n+m$ equations
are {\it not} linearly independent. When the particle numbers are fixed,
automatically the quantum numbers are fixed as well (the opposite, of
course, is not true). For example the equation for the conservation of the
baryon number can be expressed as a linear combination of equations for the
conservation of the number of certain particle species
\[
<B>-<B'>=(<N_n>-<N_{n}'>)+(<N_p>-<N_{p}'>)+\cdots
\]
\begin{equation}
\hspace{3.3cm}
-(<N_{\bar{n}}>-<N_{\bar{n}}'>)-(<N_{\bar{p}}>-<N_{\bar{p}}'>)-\cdots
\;.
\end{equation}

One might think that a way out of the problem of the linear dependency of the
equations (11) and (12) would be to reduce the number of particle number
fugacities by $n$ in which case $n$ particle species would be described by only
quantum numbers fugacities. Let us suppose for simplicity that the hadronic
gas is composed only of particles $1$ and $2$ and the only relevant quantum
number is $B$. If one decided to describe the hadronic gas with the
fugacities $\lambda_B$ and $\lambda_2$, then two equations would have to be
satisfied
\begin{equation}
<B>=<B'>\Leftrightarrow
V(Z_{H_1}(T) \lambda_B + Z_{H_2}(T) \lambda_B\lambda_2)=
V'(Z_{H_1}(T') \lambda_B' + Z_{H_2}(T') \lambda_B'\lambda_2')
\end{equation}
\begin{equation}
<N_2>=<N_2'>\Leftrightarrow
V(Z_{H_2}(T) \lambda_B\lambda_2)=
V'(Z_{H_2}(T') \lambda_B'\lambda_2')\;.
\end{equation}
Using (15) in (14) one can solve for the final $\lambda_B'$ to find
\begin{equation}
\lambda_B'=\frac{V}{V'}\lambda_B \frac{Z_{H_1}(T)}{Z_{H_1}(T')}=
\frac{VT}{V'T'}\lambda_B \frac{K_2(m_1/T)}{K_2(m_1/T')}\;.
\end{equation}

On the other hand if someone had decided to use the set of fugacities
$\lambda_B$ and $\lambda_1$ he would arrive in a similar way to the relation
\begin{equation}
\lambda_B'=\frac{V}{V'}\lambda_B \frac{Z_{H_2}(T)}{Z_{H_2}(T')}=
\frac{VT}{V'T'}\lambda_B \frac{K_2(m_2/T)}{K_2(m_2/T')}\;.
\end{equation}

It is obvious from comparing (16) and (17)\footnote{The hadron masses are
in general different ($m_1 \neq m_2$).} that the value of baryon number
fugacity depends on the choice of which particle number fugacities are
kept. This, of course, is undesirable.

Two alternative choices thereby present themselves. The first is to drop the
quantum numbers fugacities after chemical freeze-out and describe the
evolution of the system with only the particle number fugacities. The second
is to keep the quantum numbers fugacities with the ambiguity that accompanies
them. In either case the product of fugacities which accompany the part of
the partition function associated with each particle species has no problem
to be evaluated. Thus quantities like the baryon density can be calculated.

{\large{\bf Figure Captions}}
\newtheorem{g}{Figure} 
\begin{g}
\rm Contours (thick lines) that follow hadronic systems after chemical
freeze-out on ($T,n_B$) plane for 3 interactions at SPS, calculated through
FPN (model of Fixed Particle Numbers). On the same graph points
(dotted lines) calculated through an IHG model that conserve entropy and
baryon number are depicted.
\end{g}
\begin{g}
\rm The points of the IHG model of Figure 1 on the ($T,\mu_B$) plane
(dotted lines). The thick lines represent calculation through IHG of the
baryon chemical potential that leads for a given temperature to the same
baryon density as FPN.
\end{g}
\begin{g}
\rm The ratio of the volume $V'$ of the hadronic system to its volume
$V^{ch}$ at chemical freeze-out as function of temperature calculated for
the models FPN and IHG of Figure 1.
\end{g}

\vspace{0.3cm}
{\large{\bf Table Captions}} 
\newtheorem{f}{Table} 
\begin{f} 
\rm Chemical freeze-out parameters calculated for different interactions
at SPS and the corresponding references.
\end{f}
\begin{f} 
\rm Thermal freeze-out temperature calculated in two different references
for the $Pb+Pb$ interaction and the corresponding computation of baryon
density through FPN. The upper errors of baryon density correspond to the
upper errors of temperature. The same is true for the lower errors.
\end{f}

\vspace{0.3cm}
\begin{center}
\begin{tabular}{|c|ccc|c|} \hline
Experiment & $T^{ch} (MeV)$ & $\mu_B^{ch} (MeV) $ & $\gamma_s^{ch}$ &
 Reference \\ \hline\hline
S+S 200 $A\cdot GeV$ & $180.5\pm10.9$ & $220.2\pm18.0$ & $0.747\pm0.048$ & [18,19]\\
S+Ag 200 $A\cdot GeV$ & $178.9\pm8.1$ & $241.5\pm14.5$ & $0.711\pm0.063$ & [18,19]\\
Pb+Pb 158 $A\cdot GeV$ & $174.7\pm6.7$ & $240\pm14$    & $0.900\pm0.049$ & [20]\\
\hline
\end{tabular} 
\end{center} 

\begin{center}
Table 1.
\end {center} 

\vspace{0.3cm}
\begin{center}
\begin{tabular}{|c|cc|c|} \hline
Experiment & $T^{ther} (MeV)$ & Reference& $n_B^{ther}$ ($fm^{-3}$) \\
\hline\hline
Pb+Pb 158 $A\cdot GeV$ & $120\pm12$   & [21] & $0.099^{+0.022}_{-0.019}$ \\
Pb+Pb 158 $A\cdot GeV$ & $95.8\pm3.5$ & [22] & $0.0627^{+0.0047}_{-0.0045}$ \\
\hline
\end{tabular} 
\end{center} 

\begin{center}
Table 2.
\end {center}

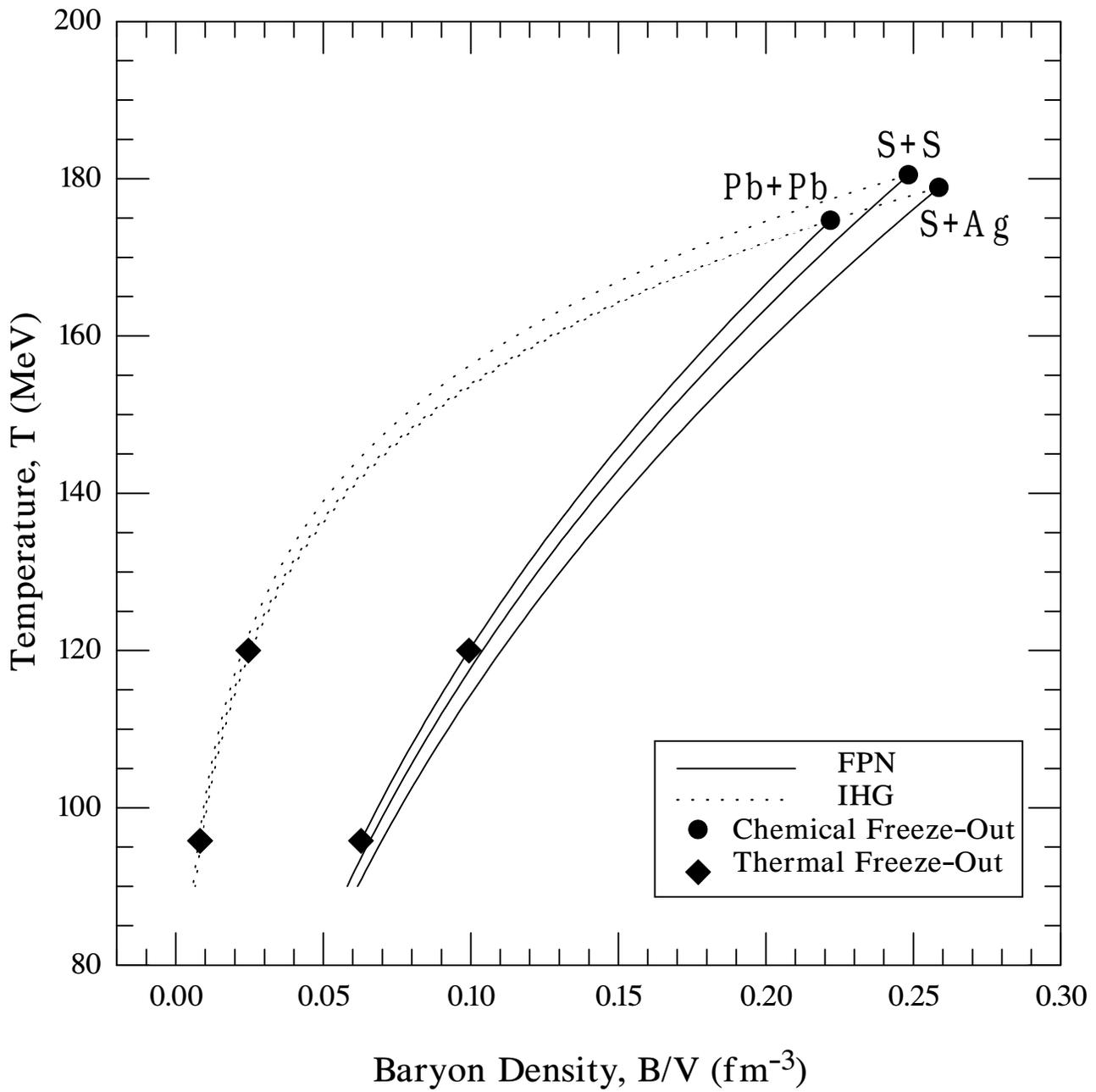
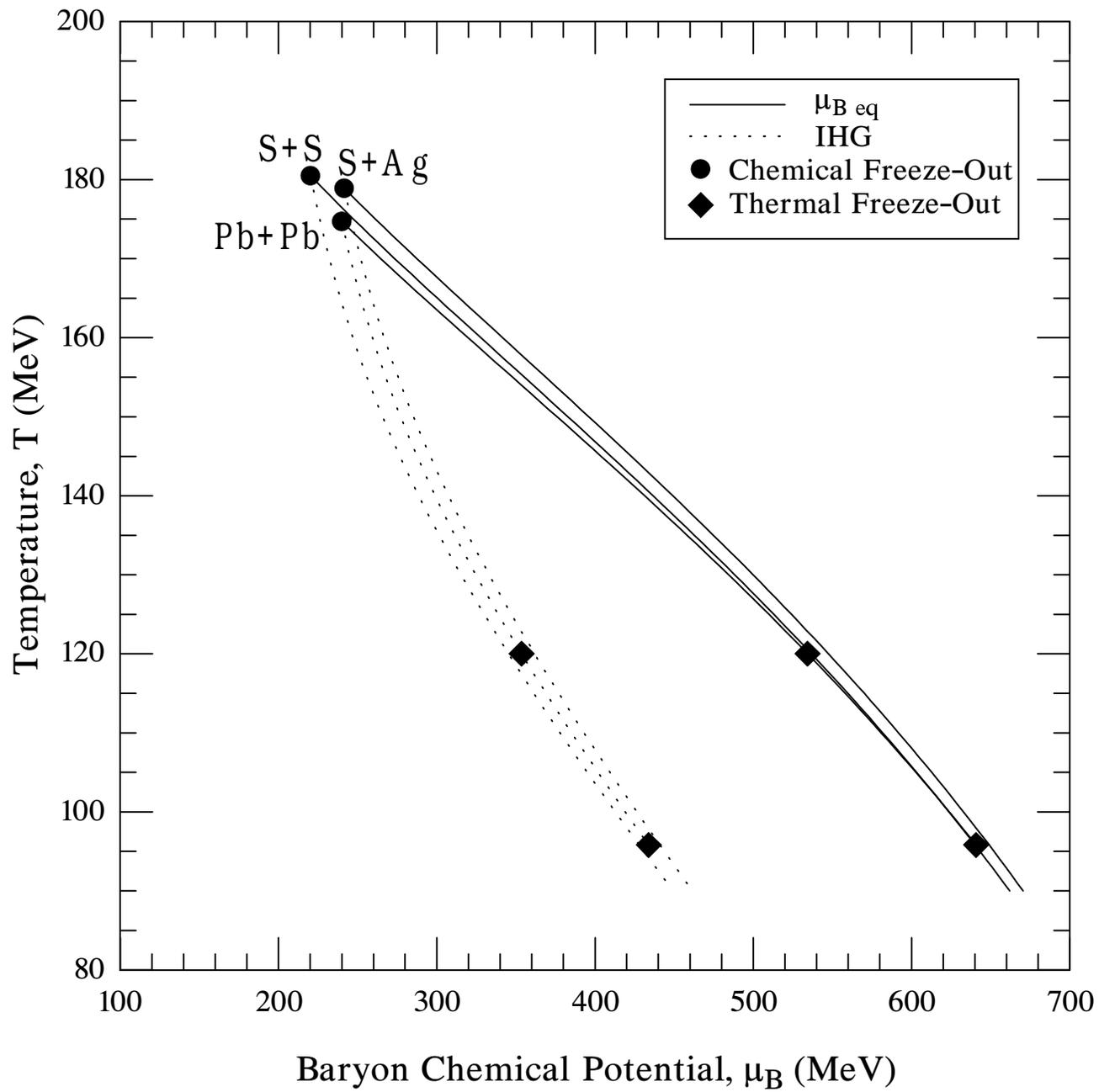
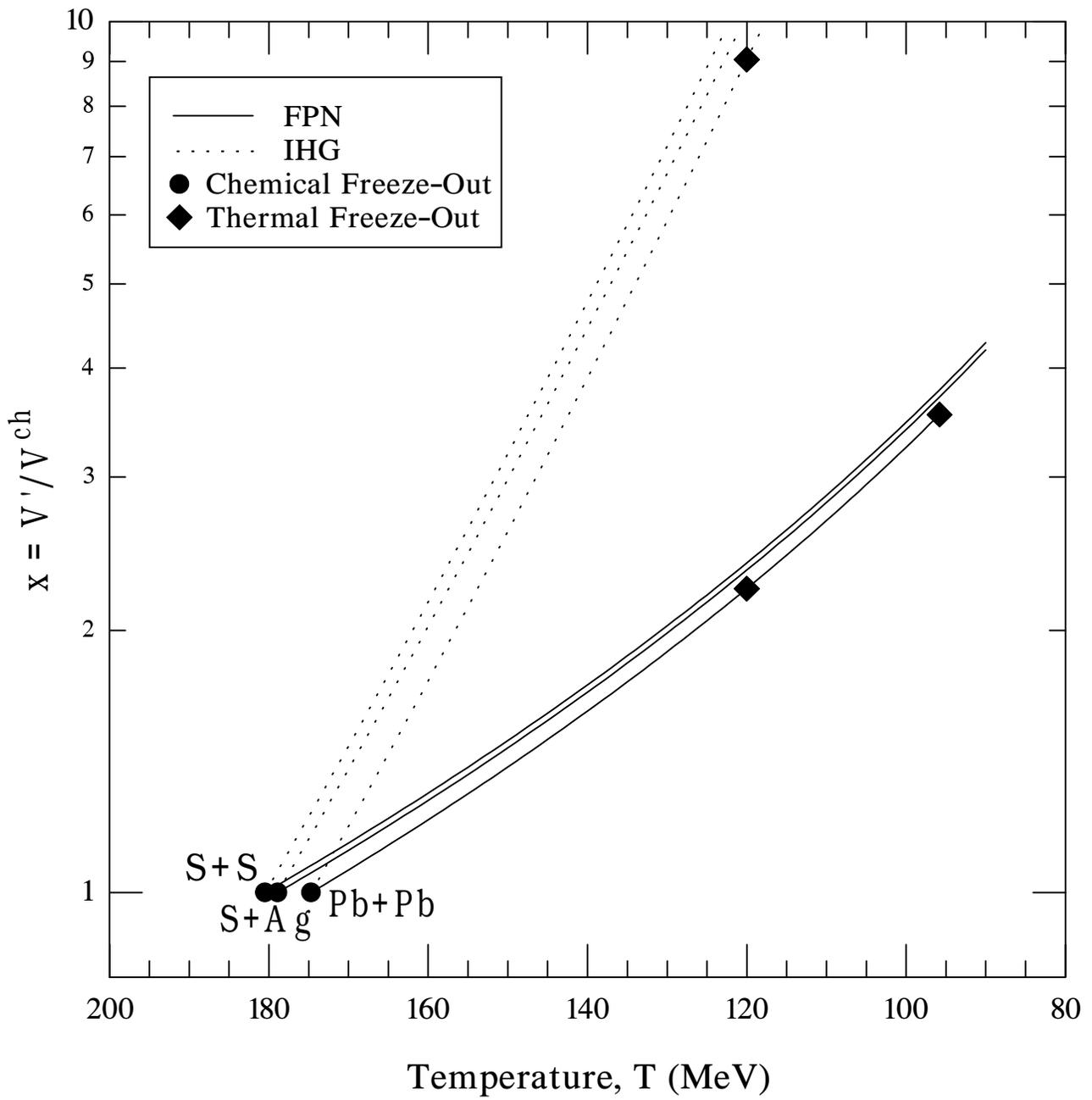
\end{document}